\documentclass[11pt]{article}

\usepackage{amsmath}  % needed for \tfrac, \bmatrix, etc.
\usepackage{amsfonts} % needed for bold Greek, Fraktur, and blackboard bold
\usepackage{amsthm,amssymb,euscript,amscd,subfigure,latexsym,psfrag, url,hyperref}
\usepackage{epstopdf,graphicx}

\begin{document}

\title{On the Self-Recovery Phenomenon in the  \\ Process of Diffusion} 
\author{Dong Eui Chang \\
Department of Applied Mathematics, \\ 
University of Waterloo, \\
Waterloo, ON N2L 3G1, Canada, \\
dechang@uwaterloo.ca\\ [2mm]
and\\ [2mm]
Soo Jeon\\
Department of Mechanical and Mechatronics Engineering, \\
University of Waterloo, \\ 
Waterloo, ON N2L 3G1, Canada \\
soojeon@uwaterloo.ca
}

\date{June 5, 2013}
\maketitle %\maketitle must follow title, authors, abstract and \pacs

\begin{abstract}
We report a new phenomenon, called self-recovery, in the process of diffusion in a region with boundary.  Suppose that a diffusing quantity is uniformly distributed initially and then gets excited by  the change in the boundary values over a time interval. When the boundary values return to their  initial values and stop varying afterwards, the value of a physical quantity related to the diffusion automatically comes back to its original value. This self-recovery phenomenon has been discovered and fairly well understood for finite-dimensional mechanical systems with viscous damping.  In this paper, we show that  it  also occurs in the process of diffusion.   Several examples are provided  from fluid flows,  quasi-static electromagnetic fields and heat conduction. In particular, our result in fluid flows provides a dynamic explanation for the famous experiment by Sir G.I. Taylor with glycerine in an annulus on kinematic reversibility of low-Reynolds-number flows.
\end{abstract}

\section{Introduction} 

\begin{figure}[!htp]
\centering
\includegraphics{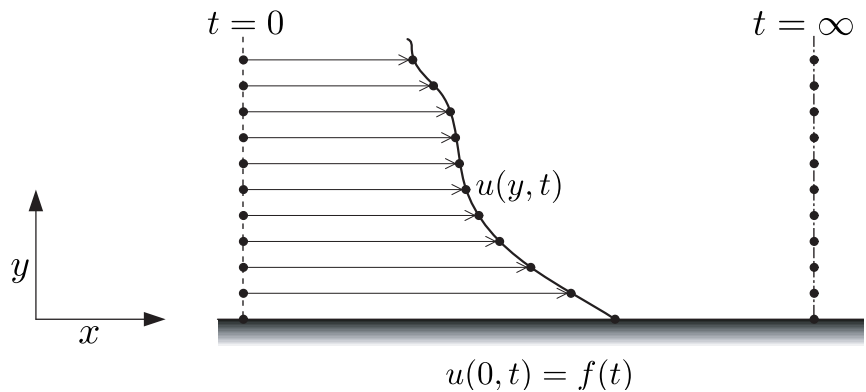}
\caption{Fluid in the space that is horizontally infinite long on both sides and vertically infinitely long upward. The bottom of the reservoir moves horizontally at the velocity of $f(t)$.}
\label{figure.Stokes1}
\end{figure}

We consider the diffusion equation
\[
\frac{\partial \phi (x,t)}{\partial t} = D\nabla^2 \phi (x,t),
\]
  where $\phi$ denotes the diffusing quantity  in a medium with boundary and $D$ the diffusion constant.  The term $D\nabla^2 \phi$ physically means dissipation that is linear in $\phi$, and the boundary value of $\phi$ specified as a function of time $t$ can be interpreted as an {\it external force} or {\it input} to this diffusing {\it system}. We shall see that the dissipation has a kind of memory. For example, let us consider a viscous flow of an incompressible fluid in the region in Fig. \ref{figure.Stokes1}. There is no pressure gradient in the region and the fluid is at rest initially. Suppose that the boundary at the bottom moves to the right by 1 m and then remains at rest. Due to the  symmetry of the configuration, the fluid flows in the $x$ direction and its velocity, which is a function of height $y$ and time $t$, obeys the diffusion equation. It turns out that every fluid particle in the region moves to the right exactly by  1 m  and comes to a stop as $t \rightarrow \infty$. Namely,  the fluid remembers its initial configuration relative to the boundary. This recovery phenomenon due to dissipation is analyzed in detail in  the present paper.

\begin{figure}[!htp]
\centering
\includegraphics{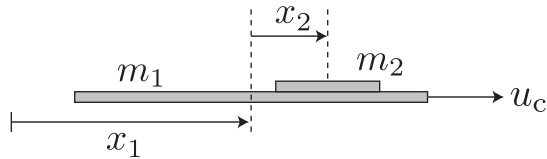}
\caption{Two plates where there is  viscous damping between the plates.}
\label{figure.two_plates}
\end{figure}

The self-recovery phenomenon induced by viscous damping friction  is well understood for finite-dimensional mechanical systems \cite{ChJe13:ASME,ChJe13:JNS}.  Hence, as motivation, we first take a simple example to recall the results in  \cite{ChJe13:ASME,ChJe13:JNS}.
Consider the system of two plates in Fig. \ref{figure.two_plates}, where  plate 1 is assumed to be much longer than  plate 2. The position of  plate 1 is denoted by $x_1$ and the relative position of  plate 2 with respect to  plate 1 by $x_2$. The mass of plate $i$ is denoted by $m_i$ for $i = 1,2$. We assume that  plate 1 is controlled by a control force $u_{\rm c}$ and that there exists a viscous damping force $-b \dot x_2$ between the plates. We assume there are no other forces on the system. Then the equations of motion are given by
\begin{align}
m_1 \ddot x_1  &= b \dot x_2 + u_{\rm c}, \label{plate:eq:1}\\
m_2 \ddot x_1 + m_2 \ddot x_2 &= -b\dot x_2.\label{plate:eq:2}
\end{align}
The two plates are aligned along their centers and both plates are at rest at $t=0$ such that
\[
x_1(0) = x_2(0) = \dot x_1(0) = \dot x_2(0) = 0.
\]
Choose a control $u_{\rm c}$ such that $\dot x_1(t)$ converges exponentially to zero as $t$ tends to infinity, which is always possible.

Integrating Eq. \eqref{plate:eq:2}, one can see that the following {\it damping-added momentum} is conserved:
\begin{equation}\label{damping:J}
J_{\rm d} := m_2 \dot x_1 (t) + m_2 \dot x_2 (t) + b x_2 (t) = 0
\end{equation}
for all $t$.
Solve Eq. \eqref{damping:J} for $x_2(t)$:
\begin{equation}
x_2(t) = - e^{-\frac{b}{m_2}t} \int_0^t e^{\frac{b}{m_2}\tau}\dot x_1 (\tau) d\tau.
\end{equation}
One can see that when  plate 1 is moving forward, i.e., $\dot x_1(t) >0$, plate 2 falls behind relative to plate 1 since $x_2(t) <0$.  However,
since $\dot x_1(t)$ decays exponentially to zero, we have
\[
\lim_{t\rightarrow\infty} x_2(t) = 0.
\]
In other words, as  plate 1 comes to a stop,  plate 2 catches up and comes back to its initial position relative to  plate 1. This phenomenon is called damping-induced self-recovery \cite{ChJe13:ASME,ChJe13:JNS}.

There is another interesting phenomenon in this system due to the viscous damping although it is not directly related to the main results of this paper. Suppose that one chooses a control $u_{\rm c}$ such that plate 1 moves at a constant speed over a time interval $[T_1, T_2]$ with $T_1 << T_2$ , i.e.,
\[
\dot x_1 (t) = c_1
\]
for all $t \in [T_1, T_2]$.  Solving Eq. \eqref{damping:J} for $x_2(t)$ and differentiating it, one can obtain
\[
\dot x_2(t) = - \frac{b}{m_2}\left (x_2(T_1) - \frac{c_1m_2}{b} \right )e^{-\frac{b}{m_2}(t-T_1)}
\]
for all $t \in [T_1, T_2]$.  Hence,  $\dot x_2(t) \approx 0$ for all sufficiently large $t$ in $[T_1, T_2]$. In other words, while  plate 1 is moving at a constant speed,  plate 2 moves at the same speed in the inertial frame. Namely,  plate 2 does not slide on  plate 1 while  plate 1 is moving at a constant speed. This phenomenon is called damping-induced boundedness \cite{ChJe13:ASME,ChJe13:JNS} since the distance $|x_2(t)|$ between  plate 1 and plate 2 does not grow indefinitely while the speed of  plate 1 is bounded.

It should be now clear by analogy  that the self-recovery phenomenon must occur in the fluid flow in Fig. \ref{figure.Stokes1}. In this paper, we provide examples of self-recovery phenomena in various diffusion processes from fluid flows, quasi-static electromagnetic fields and heat conduction. Since there are other situations  where the diffusion equation dictates the dynamics, there should be more instances of self-recovery in nature.
\section{Damping-Induced Self-Recovery in Fluid Flows}

\subsection{Modified Stokes' Problem}

Consider the  viscous flow of an incompressible fluid  in Fig. \ref{figure.Stokes1}, where an infinitely long plate is at the bottom above which  a fluid exists.
Initially, both the plate and the fluid are assumed at rest.  The boundary plate begins to move horizontally at $t=0$ and returns to rest. More precisely,  the velocity function $f(t)$, $0\leq t < \infty$,  of the boundary is such that  $f(0) = 0$ and $f(t)$ converges exponentially to zero as $t\rightarrow \infty$.\footnote{ In this paper, a function $f(t), \,\, 0 \leq t < \infty$, is said to exponentially converge to zero as $t \rightarrow \infty$ if there are numbers $\sigma >0$ and $K>0$ such that $|f(t)| \leq K e^{-\sigma t}$ for all $t\geq 0$. By this definition, any function $f(t)$ that vanishes after a finite time converges exponentially to zero as $t\rightarrow \infty$.} There is no slip at the boundary.

Due to the symmetry of the configuration, the fluid flows only in the $x$ direction. Its velocity $u(y,t)$  at height $y$ and time $t$ satisfies the Navier-Stokes equation that reduces to the diffusion equation
\[
\frac{\partial u}{\partial t} = \nu \frac{\partial^2 u}{\partial y^2}
\]
and the boundary conditions
\begin{align*}
&u(y,0) = 0,\\
& u(0,t) = f(t),\\
&u(y=\infty, t) = 0,
\end{align*}
where $\nu$ is the constant kinematic viscosity of the fluid; refer to section 4.3 of  \cite{Ba67} for derivation of the Navier-Stokes equation for this configuration.
Let $\hat f(s)$ denote the Laplace transform of $f(t)$:
\[
\hat f(s) := \int_0^\infty e^{-st} f(t)dt.
\]
By the exponential convergence assumption on $f(t)$, there is a $\sigma >0$ such that $\hat f(s)$ is analytic for all $s \in \mathbb C$ with  $\operatorname{Re}(s) > -\sigma$. Hence,
\[
\hat f(0) = \int_0^\infty f(t) dt < \infty.
\]
Let $\hat u(y,s)$ denote the Laplace transform of the solution $u(y,t)$ with respect to $t$.  Using the techniques in Chapter 7 of  \cite{Ch71},  one can obtain
\[
\hat u(y,s) = e^{-y\sqrt {s/\nu} } \hat f(s).
\]
Due to the symmetry of the configuration, all the particles at $y$ move at the same speed. So, the displacement $d(y,t)$ of each fluid particle at  $y$ over the time interval $[0,t]$  in the $x$ direction is given by
\[
d(y,t) = \int_0^t u(y,\tau ) d\tau
\]
and its Laplace transform $\hat d(y,s)$ with respect to $t$ is given by
\[
\hat d(y,s) = \frac{1}{s}e^{-y\sqrt {s/\nu} } \hat f(s).
\]
For each fixed $y$, the function $\hat d(y,s)$ has a simple  pole at $s=0$ and all  its other poles have negative real part; recall that $\hat f(s)$ is analytic for all $s\in \mathbb C$ with $\operatorname{Re}(s) > -\sigma$ for some positive number $\sigma$. Hence,
by the final-value theorem, the displacement $d(y)$  of each particle at $y$ over the time interval $[0,\infty)$ is given by
\[
d(y) = \lim_{t\rightarrow \infty} d(y,t) = \lim_{s\rightarrow 0+}s \hat d(y,s) =  \lim_{s\rightarrow 0+}e^{-y\sqrt {s/\nu} } \hat f(s) = \hat f(0) = \int_0^\infty f(t) dt.
\]
It follows that every fluid particle moves horizontally by the same amount as the displacement of the boundary over the time interval $[0,\infty)$. In other words,  each fluid particle comes back to its initial position relative to the boundary as $t \rightarrow \infty$. This is the  self-recovery phenomenon in this  fluid flow.

\subsection{Flow between Parallel Plates}
\begin{figure}[!htp]
\centering
\includegraphics{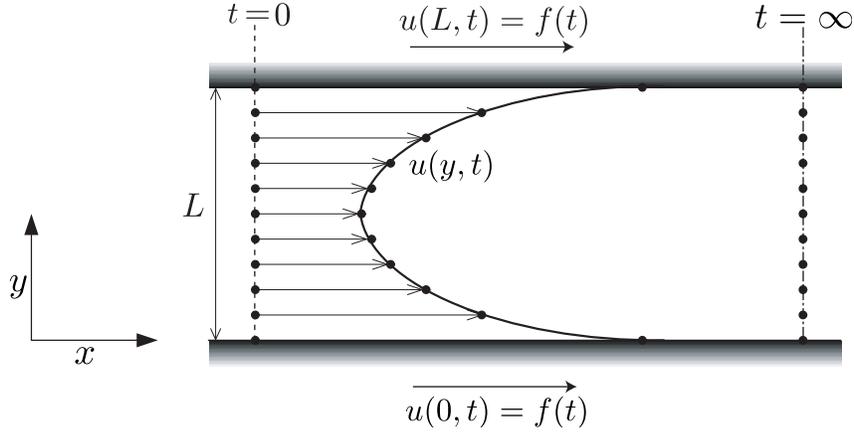}
\caption{Flow between two infinitely long parallel plates.}
\label{figure.pipe}
\end{figure}
Consider the viscous flow of an incompressible fluid between parallel plates in Fig. \ref{figure.pipe}, where both boundaries move at the same velocity $f(t)$.  Assume that the fluid is at rest initially. Also assume that $f(0)=0$ and that $f(t)$ converges exponentially to zero as $t\rightarrow \infty$. There is no slip at the boundary. Due to symmetry, the fluid flows only in the $x$ direction. Let $u(y,t)$ denote the velocity of the flow at $(y,t)$. It satisfies the Navier-Stokes equation that reduces to the diffusion equation
\[
\frac{\partial u}{\partial t} = \nu \frac{\partial^2 u}{\partial y^2}
\]
and the boundary conditions
\begin{align*}
&u(y,0) = 0,\\
& u(0,t) = f(t),\\
&u(L, t) = f(t).
\end{align*}
Let $\hat u(y,s)$ denote the Laplace transform of the solution $u(y,t)$ with respect to $t$. Using the techniques in Chapter 7 of \cite{Ch71},  one can obtain
\[
\hat u(y,s) = \frac{\sinh (y\sqrt{ s/\nu}) + \sinh ((L-y)\sqrt {s/\nu}) }{\sinh (L\sqrt{ s/\nu})} \hat f(s),
\]
where $\hat f(s)$ denotes the Laplace transform of $f(t)$.
Due to the translational symmetry in the configuration, all the particles at $y$ move at the same speed. So, the displacement $d(y,t)$ of each fluid particle at  $y$ over the time interval $[0,t]$  in the $x$ direction is given by
\[
d(y,t) = \int_0^t u(y,\tau ) d\tau.
\]
Its Laplace transform $\hat d(y,s)$ with respect to $t$ is given by
\[
\hat d(y,s) = \frac{1}{s}\cdot \frac{\sinh (y\sqrt s) + \sinh ((L-y)\sqrt s) }{\sinh (L\sqrt s)} \hat f(s).
\]
The function $\hat d(y,s)$ has a simple  pole at $s=0$ and all  its other poles have negative real part;  refer to Chapter 7 of  \cite{Ch71}. Hence,
by the final-value theorem,  the displacement $d(y)$  of each particle at $y$ over the time interval $[0,\infty)$ is computed as
\begin{align*}
d(y) &= \lim_{t\rightarrow \infty} d(y,t) \\
&= \lim_{s\rightarrow 0+}s \hat d(y,s)\\
& =  \lim_{s\rightarrow 0+}\frac{\sinh (y\sqrt{ s/\nu}) + \sinh ((L-y)\sqrt {s/\nu}) }{\sinh (L\sqrt{ s/\nu})} \hat f(s) \\
& =  \lim_{s\rightarrow 0+}\frac{\sinh (y\sqrt{ s/\nu}) + \sinh ((L-y)\sqrt {s/\nu}) }{\sinh (L\sqrt{ s/\nu})}  \cdot  \lim_{s\rightarrow 0+}\hat f(s) \\
&=  1\cdot \hat f(0) \\
&= \int_0^\infty f(t) dt,
\end{align*}
where l'H\^{o}pital's rule is used for the second last equality.
It follows that every fluid particle moves horizontally by the same amount as the displacement of the boundary over the time interval $[0,\infty)$. In other words,  each fluid particle comes back to its initial position relative to the boundary as $t \rightarrow \infty$. This is the  self-recovery phenomenon.

\paragraph{Remark.} Suppose that the lower boundary and the upper one move at different speeds, say the one at $y=0$ moves at the velocity of $f(t)$ and the one at $y=L$ at the velocity of $g(t)$. Then, $\hat u(x,s)$ is computed as
\[
\hat u(x,s) = \frac{\hat f(s) \sinh ((L-y) \sqrt{s/\nu}) + \hat g(s) \sinh (y\sqrt{s/\nu})}{\sinh (L\sqrt{s/\nu)}}.
\]
One can show that the displacement $d(y)$ of a particle at $y$ over the time interval $[0, \infty)$ is given by
\[
d(y) = \frac{L-y}{L}\int_0^\infty f(t) dt + \frac{y}{L}\int_0^\infty g(t)dt,
\]
which becomes  a constant $\int_0^\infty f$ if $\int_0^\infty f = \int_0^\infty g$.  In other words,  even if the two boundaries move differently, as long as their final displacements are equal,  all the fluid particles move by the same amount of the displacement as the boundaries.
%\end{remark}

\subsection{Flow inside a Cylinder}
\begin{figure}[!htp]
\centering
\includegraphics{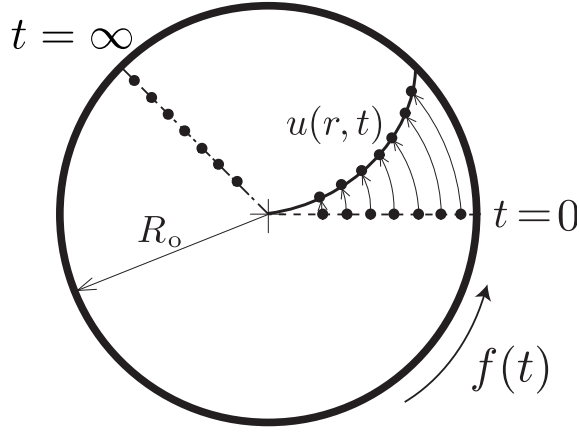}
\caption{Geometry for flow  inside a cylinder that rotates at angular velocity $f(t)$.}
\label{figure.cup}
\end{figure}
Consider the viscous flow of an incompressible  fluid inside a cylinder of radius $R_{\rm o}$ that rotates at the {\it angular velocity} $f(t)$, where $f (0)=0$  and $f (t)$ converges exponentially to zero as $t\rightarrow \infty$. Assume that the fluid is at rest initially. Due to rotational symmetry in the configuration, the fluid flows only coaxially.  Let  $u(r,t)$ denote the {\it tangential velocity} of the fluid at radius $r$ and time $t$. It satisfies the Navier-Stokes equation in  polar coordinates
\[
\frac{\partial u}{\partial t} = \nu \left ( \frac{\partial^2 u}{\partial r^2} + \frac{1}{r}\frac{\partial u}{\partial r} - \frac{u}{r^2}\right )
\]
and the boundary conditions
\begin{align*}
&u(r,0) = 0,\\
&u(0,t) = 0,\\
&u(R_{\rm o},t) = R_{\rm o}f(t).
\end{align*}
Refer to section 4.5 of  \cite{Ba67} for derivation of the Navier-Stokes equation in the above.
Let $\hat u(r,s)$ denote the Laplace transform of the solution $u(r,t)$ with respect to $t$. From section 148 of  \cite{Ch71}, it follows that
\begin{equation}\label{u:cyl}
\hat u(r,s) = \frac{J_1(i r\sqrt{s/\nu})}{J_1(i R_{\rm o}\sqrt{s/\nu})} R_{\rm o} \hat f (s),
\end{equation}
where $\hat f (s)$ is the Laplace transform of $f (t)$ and $J_1 (\cdot )$ is the Bessel function of the first kind with index 1, which is  given by
\[
J_1(t) =  \sum_{j=0}^\infty \frac{(-1)^j}{j! (j+1)!} \left ( \frac{t}{2}\right )^{2j+1}.
\]
 The angular velocity $\Omega (r,t)$ of the fluid at radius $r$ and time $t$ is given by
\[
\Omega (r,t) = \frac{1}{r} u(r,t),
\]
so its Laplace transform $\hat \Omega (r,s)$ with respect to $t$ is given by
\[
\hat \Omega (r,s) = \frac{1}{r} \hat u(r,s).
\]
Due to the rotational symmetry in the configuration, all particles at radius $r$ rotate at  the same angular velocity, so the net angle $\theta (r, t)$ swept by each particle at radius $r$ over the interval $[0,t]$ is given by
\[
\theta (r,t) = \int_0^t \Omega (r,\tau) d\tau,
\]
and its Laplace transform $\hat \theta (r,s)$ with respect to $t$ is given by
\[
\hat \theta (r,s) = \frac{1}{s}\hat \Omega (r,s) = \frac{1}{sr}\hat u(r,s).
\]
Since the Bessel function $J_1(\cdot)$ has only real roots, the function $\hat \theta (r,s)$ for each fixed $r$ has a simple pole at $s=0$ and all  its other poles have negative real part \cite{Ch71}.
By the final-value theorem the angle $\theta (r)$ swept by each particle at radius $r$ over the time interval $[0,\infty)$ is given by
\begin{align*}
\theta (r) &= \lim_{t\rightarrow\infty} \theta (r,t) \\
&=\lim_{s\rightarrow 0+}s \hat \theta (r,s)\\
&= \lim_{s\rightarrow 0+}\frac{R_{\rm o}}{r}\frac{J_1(i r\sqrt{s/\nu})}{J_1(i R_{\rm o}\sqrt{s/\nu})} \hat f (s)\\
&= \lim_{s\rightarrow 0+}\frac{R_{\rm o}}{r}\frac{J_1(i r\sqrt{s/\nu})}{J_1(i R_{\rm o}\sqrt{s/\nu})}  \cdot  \lim_{s\rightarrow 0+} \hat f (s)\\
&= 1 \cdot \hat f(0) \\
& = \int_0^\infty f(t) dt,
\end{align*}
where l'H\^{o}pital's rule is used in the second last equality.
Recall that   $\int_0^\infty f$ is the angle by which the cylinder has rotated over the time interval $[0,\infty)$.  It follows that every fluid particle asymptotically comes back to its initial position relative to the boundary  as $t\rightarrow \infty$, which is the  self-recovery in the fluid flow inside the cylinder.

\paragraph{Remarks.}  1. We can interpret the result in terms of vorticity $\omega$ of the fluid that is given by
\[
\omega(r,t)  = \frac{\partial u}{\partial r} + \frac{u}{r},
\]
which satisfies the diffusion equation
\[
\frac{\partial \omega}{\partial t} = \nu  \left ( \frac{\partial^2 \omega }{\partial r^2} + \frac{1}{r} \frac{\partial \omega }{\partial r}\right )
\]
in  polar coordinates; see Eq. (4$\cdot$5$\cdot$6) in  \cite{Ba67} for verification of this equation. By Eq. \eqref{u:cyl},   the Laplace transform $\hat \omega(r,s)$ of $\omega (r,t)$ with respect to $t$ is computed to be
\[
\hat \omega (r,s) =i \sqrt{\frac{s}{\nu}} \cdot \frac{J_1^\prime(i r\sqrt{s/\nu})}{J_1(i R_{\rm o}\sqrt{s/\nu})} R_{\rm o} \hat f (s) + \frac{1}{r}\hat u(r,s).
\]
Using the final value theorem, it is easy to show that
\[
\int_0^\infty \omega (r,t) dt = 2 \int_0^\infty f(t) dt,
\]
which is independent of radius $r$.  This is a manifestation of self-recovery in  terms of vorticity.

2. The self-recovery phenomenon in the cylindrical fluid flow can be easily demonstrated at home using a turntable. Put some honey and water into a bowl, mix them well and place the bowl at the center of the turntable. Then, drop some pieces of paper on the surface of the fluid along a radial direction.  If one turns the turntable and then stops it after  5 to 6 seconds, he can observe that the paper pieces get re-aligned in a radial direction, showing that every fluid particle goes back to its initial position in the bowl. A video of this experiment can be seen and downloaded from \cite{Youtube}.

3. From the result in this section,  we can dynamically understand the kinematic reversibility of low-Reynolds-number flows in the famous experiment  by Sir G.I. Taylor with glycerine; refer to  \cite{video:Taylor} for a film showing this experiment (from 13:13  in this film) and to \cite{notes:Taylor} for notes of this film.  In this experiment, the net number of revolutions made by the moving boundary is zero, so every fluid particle should return to its initial place as a self-recovery phenomenon.

\section{Self-Recovery in the Other Areas of Physics}

\subsection{Quasi-Static Electromagnetic Fields}  The quasi-static magnetic field ${\mathbf H}$ in a homogeneous medium with constant conductivity $\sigma$ and constant magnetic permeability $\mu$ satisfies the following diffusion equation:
\[
\frac{4\pi \mu \sigma}{c^2}\frac{\partial {\mathbf H}}{\partial t} = \nabla^2 \mathbf H,
\]
where $c$ is the speed of light; refer to Chapter VII of  \cite{LaPiLi84} for derivation of this equation. Hence, it is natural to expect a self-recovery phenomenon to occur in this situation. For example, consider a semi-infinite conducting permeable medium with a boundary as in Fig. \ref{figure.Stokes1}. Suppose that the magnetic field vector at $t=0$ is spatially constant. After subtracting the constant vector, without loss of generality we may assume that magnetic field is zero uniformly in the medium at $t=0$. The boundary surface at $y=0$ is
subject to a spatially constant but time-varying magnetic field $f(t)$ in the $x$ direction, where $f(0) =0$;     $f(t)$ converges exponentially to zero as $t\rightarrow0$;  and $f(t)$ is allowed to vary such that  the quasi-static assumption still holds.  Due to the translational symmetry of the configuration, only the $x$ component $H_x(y,t)$ of the magnetic field ${\mathbf H}$ induced inside the medium will vary and it is a function of only $y$ and $t$, i.e.
\begin{equation}\label{initial:H}
{\mathbf H} = (H_x(y,t), 0,0).
\end{equation}
   One can show that the integral of $H_x (y,t)$ over the time interval $[0,\infty)$ is given by
\begin{equation}\label{Hx:constant}
\int_0^\infty H_x(y,t) dt = \int_0^\infty f(t)dt
\end{equation}
for all $y\geq 0$, which is a constant function of $y$, manifesting a self-recovery phenomenon.  We can interpret this phenomenon in terms of charge flows.
The current density ${\mathbf J} (y,t) = \nabla \pmb{\times}  {\mathbf H}$ is given by
\[
{\mathbf J}(y,t) = \left (0,0, -\frac{\partial H_x(y,t)}{\partial y} \right ).
\]
Hence, we have
\[
\int_0^\infty {\mathbf J} (y,t) dt = \int_0^\infty  \nabla \pmb{\times} {\mathbf H} \, dt = \nabla \pmb{\times} \int_0^\infty {\mathbf H}\,  dt = 0
\]
since $ \int_0^\infty {\mathbf H} \, dt$ is a constant vector by Eqs. \eqref{initial:H} and \eqref{Hx:constant}. It follows that  the amount of net charge that has flown across a unit area  in the $z$-direction over the time interval $[0,\infty)$ is zero everywhere in the medium.

\subsection{Heat Conduction}

\begin{figure}[!htp]
\centering
\includegraphics{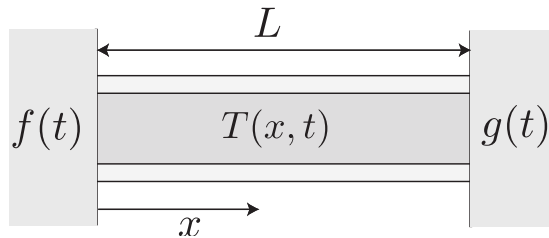}
\caption{Conduction of heat  through an insulated conducting rod, where the temperature of the left boundary varies as $f(t)$ and that of the right boundary as $g(t)$.}
\label{figure.heat}
\end{figure}
In heat conduction in a homogeneous medium, the temperature $T$ satisfies the heat equation
\[
\frac{\partial T}{\partial t } = D \nabla^2 T,
\]
where $D$ is the constant diffusion coefficient.
 Hence,  we can expect that a self-recovery phenomenon will occur in a configuration with  symmetry. Consider the situation in Fig. \ref{figure.heat}, where an insulated rod of length $L$ and sectional area $A$ is placed between two boundaries. Assume that the initial temperature is constant in the rod. By subtracting the constant value, we may then assume that  the initial temperature is uniformly zero in the rod.  There is no radiation at the boundaries. The temperature of the left boundary changes as $f(t)$ and that of the right boundary as $g(t)$, where $f(0) = g(0) = 0$ and both $f(t)$ and $g(t)$ exponentially converge to zero as $t \rightarrow \infty$ such that
\[
\int_0^\infty f(t) dt = \int_0^\infty g(t) dt.
\]
  Let $T(x,t)$ denote the temperature at $(x,t)$ in the rod. Then, the boundary conditions are given as
  \begin{align*}
  &T(x,0) = 0,\\
  &T(0,t) = f(t),\\
  &T(L,t) = g(t),
  \end{align*}
  which are the type of boundary conditions considered in Section 34 of  \cite{Ca21}.
 Solving the heat equation, one can show
 \begin{equation}\label{int:T}
\int_0^\infty T(x,t) dt = \int_0^\infty f(t)dt
 \end{equation}
for all $x$ in the rod, showing  a self-recovery phenomenon. Let us interpret this in terms of heat instead of temperature.  By the law of heat conduction,  the heat $Q(x,t)$ and the temperature $T(x,t)$ are related as
 \[
 \frac{\partial Q}{ \partial t} = -k A\frac{\partial T}{\partial x},
 \]
where $k$ is the thermal conductivity of the rod.  Hence,
 \[
Q(x, t=\infty) - Q(x,t=0) =  -\int_0^\infty  k A\frac{\partial T}{\partial x}dt = -kA \frac{\partial}{\partial x}\int_0^\infty   T\, dt= 0
 \]
 by Eq. \eqref{int:T}. It follows that the amount of net heat that has flown  across the cross section over the time integral $[0, \infty)$  is zero everywhere in the rod, which is another interpretation of the  self-recovery in heat conduction.

\section{Conclusions}
We have discovered   self-recovery phenomena in some diffusion processes in  fluid flows, quasi-static electromagnetic fields and heat conduction.  We believe that there  are more instances of self-recovery or its variants  in such areas as biology, chemistry, physics, ecology, finance and so on, where diffusion models are used.

% If you have acknowledgments, this puts in the proper section head.
%\begin{acknowledgments}
% Put your acknowledgments here.
%\end{acknowledgments}

% Create the reference section using BibTeX:
%\bibliography{your-bib-file}

\begin{thebibliography}{5}
\bibitem{ChJe13:ASME} D.E. Chang and S. Jeon, ``Damping-induced self recovery phenomenon in mechanical systems with an unactuated cyclic variable,'' {\it ASME J. Dyn. Syst. Meas. and Control}, \textbf{135} (2), 021011,  2013.

\bibitem{ChJe13:JNS} D.E. Chang and S. Jeon, ``On the damping-induced self-recovery phenomenon in mechanical systems with several unactuated cyclic variables,'' {\it  J. Nonlinear Sci.}, Submitted, {\tt arXiv:1302.2109 [math.DS]}.

%\bibitem{footnote} In this paper, a function $f(t), \,\, 0 \leq t < \infty$, is said to exponentially converge to zero as $t \rightarrow \infty$ if there are numbers $\sigma >0$ and $K>0$ such that $|f(t)| \leq K e^{-\sigma t}$ for all $t\geq 0$. By this definition, any function $f(t)$ that vanishes after a finite time converges exponentially to zero as $t\rightarrow \infty$.

\bibitem{Ba67} G.K. Batchelor, \textit{An Introduction to Fluid Dynamics}, Cambridge University Press, 1967.

\bibitem{Ca21} H.S. Carslaw, \textit{Introduction to the Mathematical Theory of the Conduction of Heat in Solids}, MacMillan and Co., Limited, London, 1921.

\bibitem{Ch71} R.V. Churchill, {\it Operational Mathematics}, 3rd Ed., McGraw-Hill, New York, 1971.

\bibitem{LaPiLi84} L.D. Landau, E.M. Lifshitz, and L.P. Pitaevskii, {\it Electrodynamics of Continuous Media}, 2nd Ed., Butterworth Heinemann, 1984.

\bibitem{Youtube} \url{http://www.youtube.com/watch?v=Z68L7amLAX8&feature=youtu.be}.

\bibitem{video:Taylor}\url{http://www.youtube.com/watch?v=51-6QCJTAjU&list=PL0EC6527BE871ABA3&index=7&feature=plpp_video}.

\bibitem{notes:Taylor}\url{http://web.mit.edu/hml/ncfmf/07LRNF.pdf}.


\end{thebibliography}

\end{document}